\documentclass[11pt,letter]{article}
\usepackage{graphicx}
\usepackage{fancybox}
\usepackage[dvips]{color}
\begin{document}
\rm

\noindent
{\Large \bf 
{Testing the Rate of False Planetary Transits 
due to Binary Star

\vspace*{2mm}\noindent Blending}}

\vspace*{8mm}

\noindent
{\Large 
G. Kov\'acs$^{1}$ \& G. Bakos$^{2,}\footnote[3]{Hubble Fellow}$}

\vspace*{5mm}

\noindent 
$^{1}$
Konkoly Observatory, Budapest, Hungary
 
\noindent
e-mail: kovacs@konkoly.hu

\vspace*{3mm}

\noindent 
$^{2}$
Harvard-Smithsonian Center for Astrophysics, Cambridge, MA, USA

\noindent
e-mail: gbakos@cfa.harvard.edu

\vspace*{10mm}
%
%
\noindent
{\Large\sf Abstract}
\vspace*{3mm}

\noindent
{\parbox {150mm}{\small
We investigate the rate of false planetary transit detection 
due to blending with eclipsing binaries. Our approach is purely 
empirical and is based on the analysis of the artificially blended 
light curves of the eclipsing binary stars in the Large Magellanic 
Cloud from the archive of the Optical Gravitational Lensing 
Experiment (OGLE). Employing parameters that characterize the 
significance of the transit and the amplitude of the variation 
out of the transit, we can substantially limit the number of 
potential false positives. Further constraint comes from the 
expected length of the transit by a possible planetary companion. 
By the application of these criteria we are left only with 18 
candidates from the full sample of 2495 stars. Visual inspection 
of these remaining variables eliminates all of them for obvious 
reasons (e.g., for visible fingerprints of orbital eccentricity). 
We draw the attention to the short-period stars, where the false 
alarm rate is especially low.}}

\vspace*{10mm}
%
%
\noindent
{\Large\sf 1. Introduction}
\vspace*{2mm}

\noindent
It is well known that blended binaries are the most significant 
sources of confusion in searching for planetary transits (Brown 2003). 
Although the ultimate approach to this problem cannot avoid the 
tedious procedure of high-dispersion spectroscopic analysis 
(e.g., Mandushev et al.~2005; Bouchy et al.~2005), there have been 
additional simple methods suggested, based purely on the analyses 
of the light curves. Drake (2003) drew the attention to the 
gravitational and thermal effects of a stellar companion on the 
luminosity variation out of the eclipse (see also Sirko \& Paczy\'nski 
2003). Seager \& Mall\'en-Ornelas (2003) suggested a careful analysis 
of the transit shape. A similar approach was recommended by Tingley 
(2004). Recently, Tingley \& Sackett (2005) studied the transit 
duration and depth as useful parameters in filtering out most of 
the stellar companion candidates. 

The goal of the present work is to apply a set of conditions, similar 
to the ones mentioned above, on a homogeneous sample of observed 
binary stars with artificial blending added, and investigate if it 
is possible to filter them out 
as false positives from a list of transit candidates.    

The data utilized in this work comes from the OGLE LMC binary 
database of Wyrzykowski et al.~(2003). It contains 2681 objects, 
but only 2495 are used (the remaining 186 stars have less than 
100 data points per star and often show high noise). There are 
EA, EB and EW stars in the sample.

\vspace*{8mm}
%
%
\noindent
{\Large\sf 2. Methodology and Tools}
\vspace*{2mm}

\noindent
For the sake of simulating realistic blending scenario, we add 
{\it constant} intensity to each light curve and adjust it to 
get nearly the same total range of variation of $0.01$~mag for 
all stars. In this way the signal-to-noise ratio remains the 
same as in the original data. The result can be extrapolated to 
higher noise level by proper scaling with the number of data 
points (see also Sect.~5). 

\noindent
The most important ingredients of our analysis are the following:
\begin{itemize}
\item 
All blended light curves are searched for periodic transits by the 
BLS method (see Kov\'acs et al.~2002). We note that the significance 
of the detection for all variables is very high.  

\item
In order to characterize the statistical significance of the 
transit in the folded light curve derived from the preceding 
analysis, we introduce the Dip Significance Parameter (DSP) 
defined as follows:
%
%
{\large
\begin{eqnarray}
{\rm DSP} & = & \delta (\sigma^2/N_{\rm tr}+A^2_{\rm OOTV})^{-{1\over 2}}
\hskip 2mm ,
\end{eqnarray}}

\vspace*{-5mm}
\noindent
where $\delta$ is the depth of the transit, $\sigma$ is the standard 
deviation of the $N_{\rm tr}$ in-transit data points, $A_{\rm OOTV}$ 
is the peak amplitude in the Fourier spectrum of the Out of Transit 
Variation (OOTV, given by the folded time series with the exclusion 
of the transit).

\item 
The significance of the main periodic signal in the OOTV is 
characterized by the Signal-to-Noise Ratio (SNR) of the Fourier 
spectrum of the OOTV (excluding the transit):
%
%
{\large
\begin{eqnarray}
{\rm SNR}_{\rm OOTV} & = & \sigma_{\rm A}^{-1}(A_{\rm OOTV}-\langle A \rangle)
\hskip 2mm ,
\end{eqnarray}}

\vspace*{-5mm}
\noindent
where $\langle A \rangle$ and  $\sigma_{\rm A}$, respectively, denote 
the average and the standard deviation of the Fourier spectrum.

\item
We would like to exclude stellar companions characterized by too 
long transit/eclipse durations. For this goal we compute the 
expected transit length (i.e., an upper limit of it) for a hypothetical 
planetary companion with G--M primaries. The following relations 
are used:
%
%
{\large
\begin{eqnarray}
a  = \left({GMP^2\over{4\pi^2}}\right)^{1\over 3}, \hskip 10mm 
tg \ \alpha = {R\over{\sqrt{a^2-R^2}}}, \hskip 10mm 
Q_{\rm tran} \equiv T_{\rm transit}/P = \alpha/\pi \hskip 2mm , 
\end{eqnarray}}

\vspace*{-3mm}
\noindent
with $a$, $R$, $M$, $P$, $G$ denoting the semi-major axis, stellar 
radius, stellar mass, orbital period and gravitational constant, 
respectively.  The $R, M$ values correspond to Main Sequence stars 
as given in Lang (1992).
\end{itemize}

\noindent
We add the following comments to the above definitions:

\noindent
Eq.~(1) takes into consideration a possible OOTV, Eq.~(2) yields 
significance limits on the OOTV. Although the two quantities are 
not completely independent, they are useful in characterizing different 
aspects of the morphology of the light curve. Eqs.~(1) and (2) 
yield more general quantification of the OOTV than a simple 
few-component Fourier fitting, assuming tidal and/or 
reflection effects (Drake 2003; Sirko \& Paczy\'nski 2003). 
Eq.~(3) assumes central transit, thereby giving an upper limit 
on the relative transit duration. More sophisticated treatment 
of the fractional transit length $Q_{\rm tran}$ can be found in, 
e.g., Tingley \& Sackett (2005). Theoretical OOTV amplitudes are 
given in Drake (2003). Methods on massive light curve analysis 
and binary model fitting have recently been published by Devor (2005).

\vspace*{7mm}
%
%
\noindent
{\Large\sf 3. Confidence levels}
\vspace*{3mm}

\noindent
In order to estimate proper cutoff values for DSP and 
${\rm SNR}_{\rm OOTV}$ for selecting significant transit-like 
events, we generate artificial time series without a real signal, 
but with a Gaussian flux distribution on the time base of the 
observed data. For testing DSP, we use all data points, whereas 
for ${\rm SNR}_{\rm OOTV}$ we exclude the observed time moments 
of transit as derived by the BLS routine on the original data. 
This test yields estimates on the low boundaries of the 
confidence levels, because the real data are correlated at 
various degrees due to the presence of the systematics 
inherent in the data reduction (see Kruszewski \& Semeniuk 2003).   

The distribution functions are shown in Fig.~1. Numerical values 
at the tails of the distributions are as follows: For DSP the 
$P<0.01$ and $P<0.001$ confidence limits are $DSP>6.4$ and $DSP>6.9$, 
respectively. The same limits for ${\rm SNR}_{\rm OOTV}$ are 
$SNR>5.5$ and $SNR>6.6$. Obviously, feasible transit candidates 
are those events that show high DSP, but ${\rm SNR}_{\rm OOTV}$ 
is low. The above values justify the use of our ``soft'' cutoffs 
(meaning that we do not even loose marginal candidates) of 
${\rm DSP}>6.0$ and ${\rm SNR}_{\rm ootv}<7.0$ for transit 
selection in this sample. 

%
%
   \begin{figure}[h]
   \centering
   \includegraphics[width=110mm]{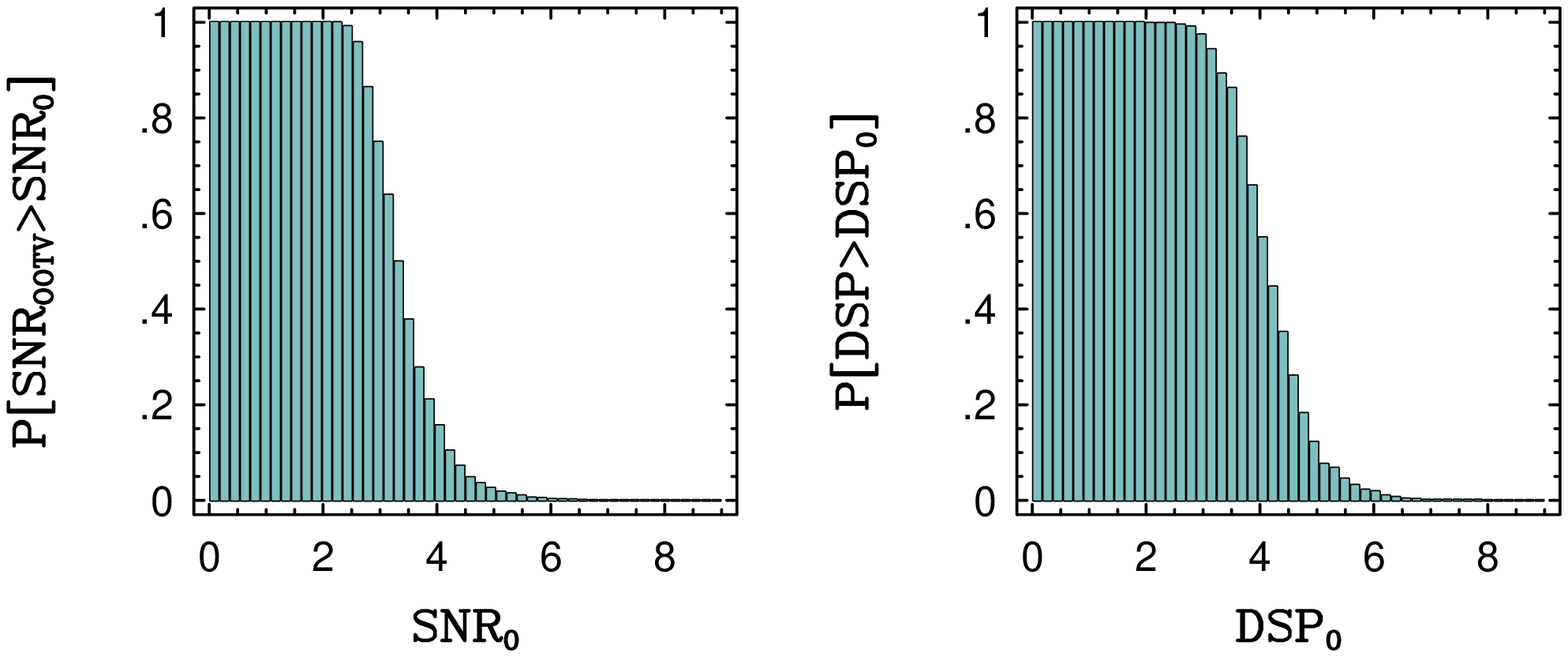}
   \flushleft
\parbox{160mm}{{\bf Fig.~1.\ }\small
      Probability distribution functions of the SNR of the OOTV 
      and of the DSP of the transit for pure Gaussian {\bf test} 
      signals generated on the OGLE timebase. These diagrams 
      yield significance levels for the above parameters when 
      employed on observed data.} 
   \end{figure}

\vspace*{5mm}
%
%
\noindent
{\Large\sf 4. Results}
\vspace*{0mm}

\vspace*{5mm}
\noindent
{\large\sl 4.1 Distribution of the OOTV peak frequencies}
\vspace*{2mm}

\noindent
Fig.~2 shows that most of the peak frequencies of the Fourier 
spectra of the OOTVs are grouped around integer frequencies 
(in the units of the orbital frequency), indicating that tidal 
and/or reflection effects are the dominating factors in causing 
OOTV. Some 75\% of the stars exhibit OOTV with peak frequencies 
$n\pm0.2$.
%
%
   \begin{figure}[h]
   \flushleft
   \includegraphics[width=80mm]{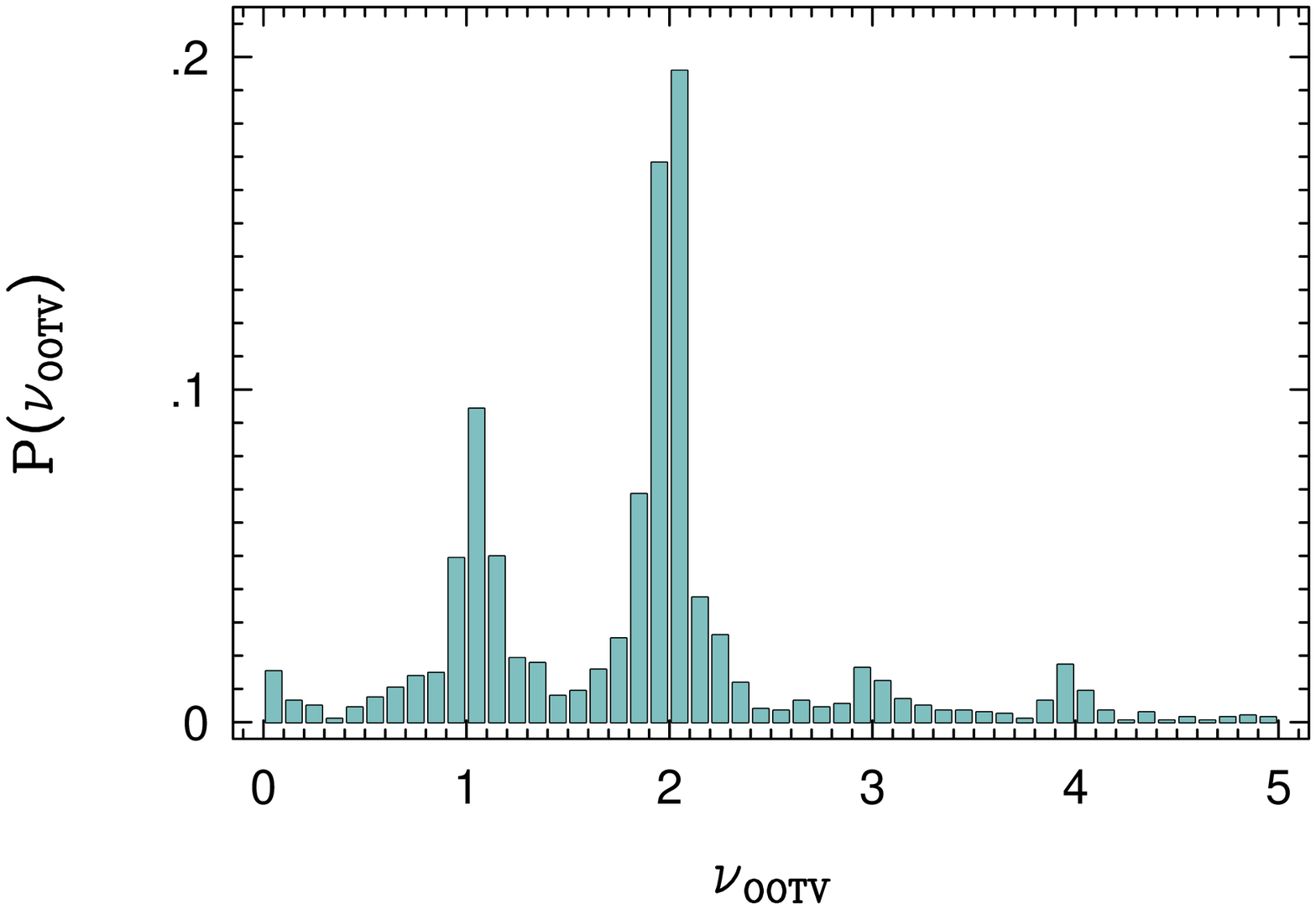}
   \flushright 
   \vspace*{-58mm}
\parbox{70mm}{{\bf Fig.~2.\ }\small
      Empirical probability density function of the 
      {\bf observed} OOTV peak frequency $\nu_{\rm OOTV}$ 
      (in the units of the orbital, i.e.~BLS peak frequency).} 
   \vspace*{40mm}
   \end{figure}

\vspace*{0mm}
\noindent
{\large\sl 4.2 Orbital frequency vs. $Q_{\rm tran}$}
\vspace*{2mm}

\noindent
The left panel in Fig.~3 shows the $Q_{\rm tran}$ values 
derived for the full sample of 2495 stars without applying any 
parameter cuts. The right panel shows the result after the 
application of the DSP and ${\rm SNR}_{\rm OOTV}$ cutoffs. 
There remain only 18 stars satisfying both of these cuts and 
the $Q_{\rm tran}<1.1Q_{\rm tran,G0V}$ constraint, 
where $Q_{\rm tran,G0V}$ is the estimated fractional transit time with a G0V primary.
(The factor $1.1$ is used for rough error allowance in the 
$Q_{\rm tran}$ values).

%
%
   \begin{figure}[h]
   \flushleft
   \includegraphics[width=77mm]{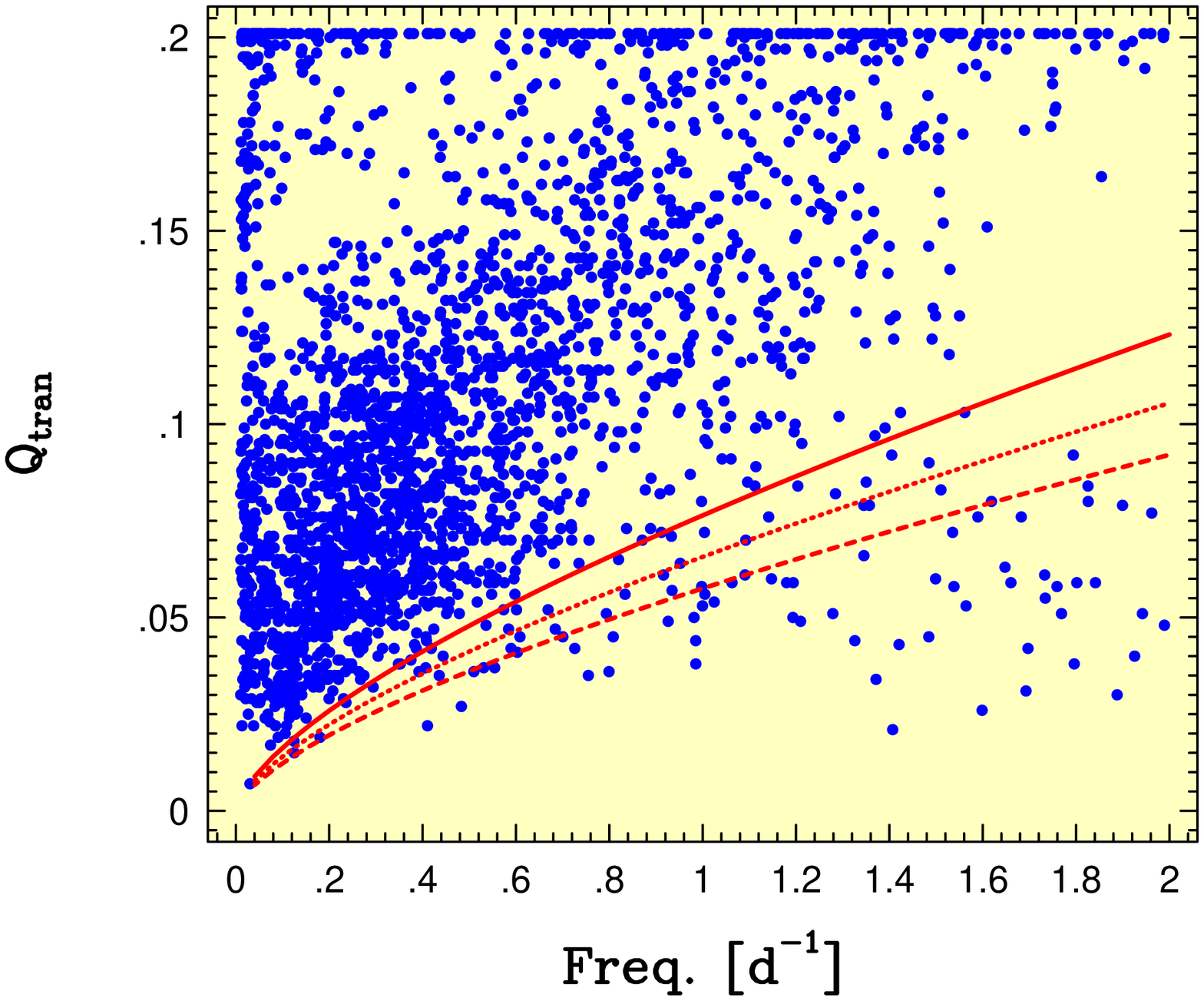}
   \flushright \vspace*{-69mm}
   \includegraphics[width=77mm]{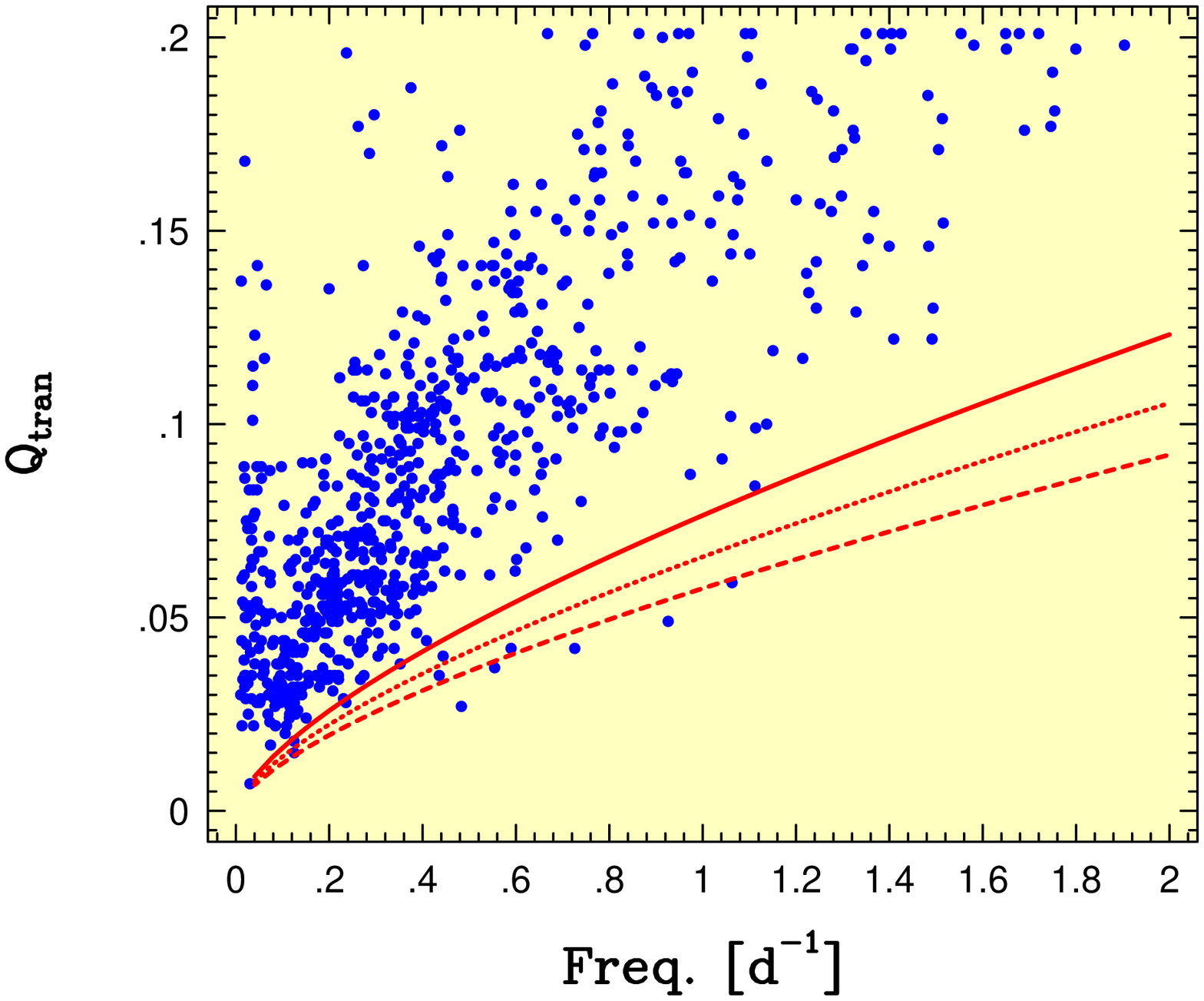}
   \flushleft
\parbox{160mm}{{\bf Fig.~3.\ }\small
      \underline{Left panel:} Fractional transit length vs. orbital 
      frequency for the sample of 2495 stars from the OGLE LMC binary 
      database. Variables with $Q_{\rm tran}>0.2$ are plotted at the 
      limiting plotting range. Continuous, dotted and dashed lines 
      denote the theoretical $Q_{\rm tran}$ values (upper limits) 
      for planetary companions with G0, M0 and K0 Main Sequence 
      primaries, respectively. \underline{Right panel:} As in the 
      panel to the left, but after applying the DSP$>6.0$ and 
      ${\rm SNR}_{\rm OOTV}<7.0$ cutoffs as discussed in the text.
}    
   \end{figure}
\newpage
\vspace*{3mm}
\noindent
{\large\sl 4.3 The `transit candidates'}
\vspace*{2mm}

%
%
%
   \begin{figure}[h]
   \centering
   \includegraphics[width=100mm]{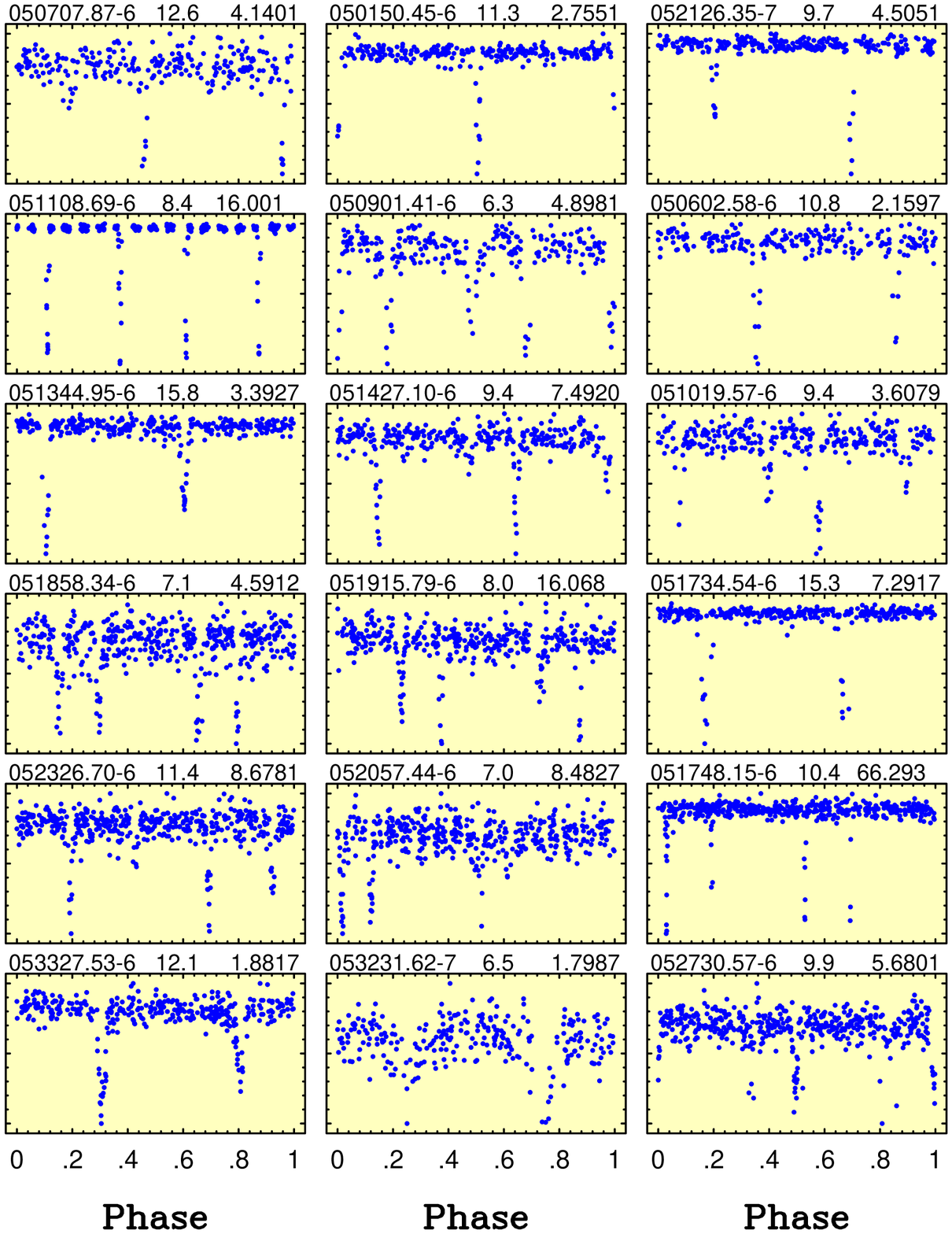}
   \flushleft
\parbox{160mm}{{\bf Fig.~4.\ }
      Folded light curves of the 18 variables 
      satisfying transit selection conditions on DSP, 
      ${\rm SNR}_{\rm OOTV}$ and $Q_{\rm tran}$ 
      (see text for details). \underline {Headers:} shortened OGLE ID, 
      DSP and folding period (in [days], {\bf twice} of the BLS period). 
      All light curves have been normalized to the same total range 
      of $0.01$~mag.
      } 
   \end{figure}

\noindent
The folded light curves of the 18 binaries that passed the 
basic steps of transit selection are shown in Fig.~4. Except 
for the following 3 stars, all exhibit obvious signs of stellar 
binary components (more or less well-defined secondary eclipse, 
uneven distribution of the eclipses due to eccentric orbit). 
Closer inspection of the remaining 3 stars shows the following:

\begin{itemize}
\item{}
{\bf 052730.57-695:} The secondary eclipse preceding the primary 
is only marginally visible (it is somewhat more easily identified 
in the 1P-folded light curve).
\item{}
{\bf 051734.54-692:} Observed in fields \#7 and 8 (field \#8  
data are shown). The light curve from field \#7 contains more data 
and yields four times longer period than the one in field \#8. 
The 4P-folded light curve shows two clear eclipses of different depth.

\item{} 
{\bf 051108.69-691:} Period is almost exactly 8~days. Folding with 
P/2 definitely shows eccentricity, as the two minima are offset 
from each other in phase.
\end{itemize}

%
%
\vspace*{5mm}
\noindent
{\Large\sf 5. Conclusions}
\vspace*{2mm}

\begin{itemize}
\item
Detailed light curve analysis combined with constraints posed by 
theoretical transit lengths shows that in a hypothetical blending 
scenario NONE of the 2495 binaries in the OGLE LMC 
database could be false positive planetary candidates. 
\item
Chance of confusion due to blending is especially low among 
short-period ($P<1.5$~d) binaries.
\item
Since for the same signal the signal-to-noise ratio scales with 
the square of the standard deviation ($\sigma$) of the noise and 
the present OGLE sample has fairly low $\sigma$, in current wide 
field surveys one needs to gather a large number of data points 
per star in order to reach the above level of confidence in selecting 
false positives.  

\item
This test indicates that careful analysis of the light curves 
(if data quality permits) indeed leads to the elimination of 
large number of binary blends, thereby narrowing down the list 
of potential planetary candidates.
\end{itemize}

\vskip-5mm
\vspace*{0mm}
\noindent
{\Large\sf Acknowledgments}
\vspace*{2mm}

\noindent
This work has been supported by the Hungarian Scientific Research Fund
(OTKA T-038437). Work of G.~B. was supported by NASA through
Hubble Fellowship grant HF-01170.01 awarded by the STScI, which is
operated by the AURA, Inc., for NASA, under contract NAS 5-26555.

%
%
\vspace*{5mm}
\noindent
{\large\sf References}
\vspace*{2mm}

\small
\noindent
Bouchy, F., Pont, F., Melo, C. et al., 2005, {\it A\&A}, {\bf 431}, 1105  

\noindent
Brown, T. M., 2003, {\it ApJ}, {\bf 593}, L125

\noindent
Devor, J., 2005, {\it astro-ph/0504399}

\noindent
Drake, A. J., 2003, {\it ApJ}, {\bf 589}, 1020 

\noindent
Kov\'acs, G., Zucker, S. \& Mazeh, T. 2002, {\it A\&A},
{\bf 391}, 369

\noindent
Kruszewski A., Semeniuk I., 2003, {\it Acta Astr.}, {\bf 53}, 241

\noindent
Lang, K. R., 1992, {\it Astrophysical Data: Planets and Stars}, 
Springer-Verlag, p. 132

\noindent
Mandushev, G. Torres, G., Latham, D. W. et al., 2005, {\it ApJ}, 
{\bf 621}, 1061

\noindent
Seager, S. \& Mall\'en-Ornelas, G., 2003, {\it ApJ}, {\bf 585}, 1038

\noindent
Sirko, E. \& Paczy\'nski, B., 2003, {\it ApJ}, {\bf 592}, 1217 

\noindent
Tingley, B., 2004, {\it A\&A}, {\bf 425}, 1125

\noindent
Tingley, B. \& Sackett, P. D., 2005, {\it astro-ph/0503575}  

\noindent
Wyrzykowski, L., Udalski, A., Kubiak, M., et al., 2003, 
{\it Acta Astr.}, {\bf 53}, 1

\end{document}